\documentclass[letterpaper, 10 pt, conference]{ieeeconf}  

\IEEEoverridecommandlockouts                              

\let\labelindent\relax
\usepackage{lmodern}
\usepackage{float}
\usepackage{graphicx}
\usepackage{amssymb,amsfonts}
\usepackage{times}
\usepackage{hyperref}
\usepackage{subcaption}
\usepackage{epstopdf}
\usepackage{color}
\usepackage{amsmath}
\usepackage{lscape}
\usepackage{enumitem}

\title{\LARGE \bf
Evaluation of Embedded Platforms for Lower Limb Prosthesis \\ with Visual Sensing Capabilities}

\author{Rafael L. da Silva$^{1}$, Nathan Starliper$^{1}$, Boxuan Zhong$^{1}$, He Helen Huang$^{2}$ and Edgar Lobaton$^{1}$
\thanks{*This work is supported by the National Science Foundation (NSF) under award CNS-1552828 and IIS-1361549.}
\thanks{$^{1}$Rafael L. da Silva, Nathan Starliper, Boxuan Zhong and Edgar Lobaton are with the Department of Electrical  and Computer Engineering, North Carolina State University, Raleigh, NC 27603, USA, emails:
        {\tt\small rdasilv2@ncsu.edu, nstarli@ncsu.edu, bzhong2@ncsu.edu, edgar.lobaton@ncsu.edu}. $^{2}$ H. Huang is within the Joint Department of Biomedical Engineering of University of North Carolina and North Carolina State University, email: {\tt\small hhuang11@ncsu.edu}}.%
}

\begin{document}

\maketitle
\thispagestyle{empty}
\pagestyle{empty}

\begin{abstract}

Lower limb prosthesis can benefit from embedded systems capable of applying computer vision techniques to enhance autonomous control and context awareness for intelligent decision making. In order to fill in the gap of current literature of autonomous systems for prosthetic legs employing computer vision methods, we evaluate the performance capabilities of two off-the-shelf platforms, the Jetson TX2 and Raspberry Pi 3, by assessing their CPU load, memory usage, run time and classification accuracy for different image sizes and widespread computer vision algorithms. We make use of a dataset that we collected for terrain recognition using images from a camera mounted on a leg, which would enable context-awareness for lower limb prosthesis. We show that, given reasonably large images and an appropriate frame selection method, it is possible to identify the terrain that a subject is going to step on, with the possibility of reconstructing the surface and obtaining its inclination. This is part of a proposed system equipped with an embedded camera and inertial measurement unit to recognize different types of terrain and estimate the slope of the surface in front of the subject.
\end{abstract}


\section{INTRODUCTION}

The relevance of providing a better quality of life to people with limb loss has only increased over time. The prospective amputee population growth is estimated to reach about $3.6$ million by 2050 in the United States alone \cite{APMR}. In order to improve amputee's quality of life, diverse types of prostheses have been employed; however, there is no prosthetic solution to date that completely addresses the lack of comfort, reliability, and safety to the user, which are claimed by several amputees as  the main reasons not to wear them \cite{ACA2}. One of the main challenges is to provide devices that are able to be natural extensions of the human body, which is an open research topic that branches across several areas: orthopedics, cardiology, neurology and many areas of engineering.

Passive prosthetic devices do not offer a high level of safety and responsiveness, frequently causing pain and overload to other limbs in order to compensate for the mismatching between body dynamics and the device itself \cite{ScienceNews}. However, active prosthetic devices have also been developed. Commercial active prostheses are controlled autonomously by a finite-state machine (FSM) \cite{Sup2008,Samuel2008}. The controller adjusts the impedance or position of the prosthetic joints based on gait phase and the user’s locomotor task (e.g. level walking, stair ascent and descent). The autonomous control cannot predict the user's intent in terms of locomotor tasks and therefore cannot adapt to walking terrains. One solution is to build a neural-machine interface that predict the user’s intended locomotor task for prosthesis operation.  Several groups have developed neural-machine interfaces (NMIs) for artificial legs to decipher neuromuscular signals (i.e. electromyographic (EMG) signals) and identify the user's intended locomotor tasks and transitions \cite{Huang2009,Huang2011,Farrell2011,Hargrove2013,Miller2013,Chen2013}. These approaches have been demonstrated to enable lower limb amputees to drive their prosthesis intuitively and switch locomotor tasks seamlessly. It has been shown that active prostheses have the potential to offer more comfort and less energy waste to the amputee compared to passive ones, \cite{Lipschutz2009,Jarasch2002,Baxter1996}. The development of intelligent prosthetic devices has been done for a while as reported by \cite{Wirta1978}. Early investigations such as \cite{Lipschutz2009, Donath1977,VanderPerre1992}, implemented control and signal processing techniques to electromechanical systems in order to provide reliable systems that bring more comfort to the amputee. This type of research is still ongoing. Contemporary approaches make use of learning algorithms applied on exoskeletons and robotic limbs \cite{JohnHopkins} with neural interfaces to the user for enhanced control, \cite{Tenore2008, Mastinu2017, Liu2014,ZhangXiao2011,DiSanto2011,Huang2010} and response to human stimuli and intent, the so-called volitional control. For example, in order to give context awareness to the volitional control system for a prosthetic leg, \cite{Liu2015} combined an inertial measurement unit (IMU) and a laser distance meter to reconstruct the terrain type and slope in front of the user \cite{Liu2015}. Lately, context awareness strategies have gained an extra resource, the use of computer vision (CV). For example, deep learning techniques have been applied to prosthetic arms in order to enhance the grasp perception, independently of object forms or size \cite{Ghazaei2017}. However, the requirements and constraints to upper limbs prosthetics are very different than the ones to lower limbs, specifically in terms of response time and robustness.

Some research has also been done related to lower limb prostheses, such as \cite{Krausz2015} have developed a depth sensing method with the help of CV. Through the use of a depth camera, they were able to recognize different types of stairs, estimating their height, depth and number of steps. Depth sensing has also been applied with the help of support vector machines (SVM) with a cubic kernel to identify different types of activities such as standing, walking, running,  and going up or down stairs \cite{Varol2016}.

\begin{figure*}[!t]
\centering
\includegraphics[width=\linewidth]{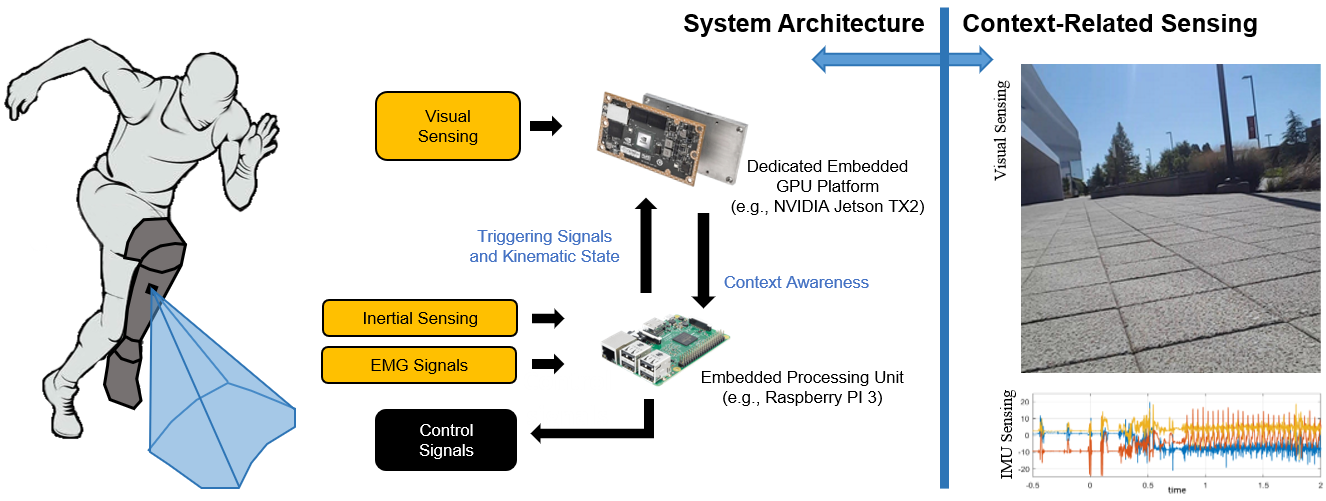}
\caption{System Overview (left) and Sensing Modalities (right).The diagram illustrates the proposed architecture for enabling visual sensing on a lower limb prosthesis. Due to real-time requirements for control, a hierarchical infrastructure is proposed where a micro-controller handles processing of fast low-bandwidth signals, and an GPU-enabled embedded platform handles the visual context awareness.} 
\label{fig:SystemOverview}
\end{figure*}

Embedding the CV methods is a critical point for an active prosthesis design. The required computational power is a major concern not only to perform activities such as feature extraction and classification but also to save power for longer operation. From the mentioned approaches, \cite{Varol2016,Krausz2015,Ghazaei2017} have applied their methods to real-time processing but carried on in a laptop with no critical constraints in terms of CPU bandwidth and power consumption. On the other hand, \cite{Tenore2008} developed the embedded CV method supported by a  digital signal processor (DSP) and an analog to digital converter (ADC) of 16,000 samples per second to perform high filtering, feature extraction, multiplications and calculation of a max function. The work shown by \cite{Mastinu2017} employed a myoelectric pattern recognition system (MPR) with low power micro-controller unit (MCU) of 32 bits with an embedded floating point unit based on ARM Cortex-M4F. The MCU was utilized to perform pattern recognition. The system developed in \cite{Liu2014} made use of two evaluation boards, one with a 16-bit ADC and another with a Digital to Analog Converter (DAC). The data processing and pattern recognition were being executed on an ARM Cortex-A8 processor.

A number of works have been done towards evaluating and comparing the computational complexity, performance, and/or power consumption of different CV algorithms on various platforms. For example, \cite{huang2017speed} studied the trade-off of speed and accuracy of different object detection algorithms; \cite{battiato2012performances} tested  classic CV tasks on mobile devices. Vision algorithms were also evaluated on embedded platforms based on specific metrics developed for different applications (e.g. unmanned aerial vehicles \cite{hulens2015choose}, driver assistance systems \cite{velez2015embedded} and automotive applications \cite{kubinger2005platform}). Other benchmark comparison were also presented by \cite{Pea2010BenchmarkingOC, Zhang2018pCAMPPC, 8336575}.

In this work, we evaluate two embedded platforms,  the NVIDIA Jetson TX2 and the Raspberry PI 3 on a dataset that we collected for context awareness for a lower limb prosthesis. The Jetson TX2 is an embedded system-on-module (SoM) with dual-core NVIDIA Denver2 + quad-core ARM Cortex-A57, 8GB 128-bit LPDDR4 and integrated 256-core Pascal GPU \cite{Jetson_TX2}. The Jetson platform has been used in many applications in robotics specifically for implementing real-time CV algorithms in autonomous vehicles and drones. The Jetson offers the option to use the OpenCV4Tegra library which is a computer vision library utilizing GPU optimized OpenCV \cite{OpenCV33} functions. GPUs have been shown to offer drastically improved computational speeds for CV and deep learning tasks. The on-board GPU offers the potential to utilize other GPU optimized deep learning libraries in future work. The Raspberry Pi 3 is an embedded system-on-chip (SoC) with a quad-core ARM Cortex-A53 and 1GB LPDDR2. It is one of the most widely used commercially available embedded platforms, used in many diverse applications for real-time control. Furthermore, we propose a hierarchical architecture (see Figure \ref{fig:SystemOverview}) that combines both platforms in order to ensure efficient real-time operation and reduced power consumption. The rest of the paper is organized as follows, section \ref{sec:motapp} describes the motivation behind our system architecture choice, section \ref{sec:plateval} explains how we are going to evaluate the embedded platforms and based on the the obtained result we conceptualize our proposed system architecture in section \ref{sec:proparch}.

\section{MOTIVATING APPLICATION}
\label{sec:motapp}

\begin{figure}[!b]
\centering
\includegraphics[width=\linewidth]{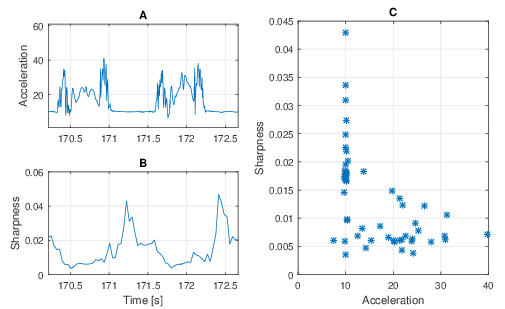}
\caption{Relationship between acceleration and sharpness. Waveforms for the (a) magnitude of acceleration and (b) sharpness. (c) Scatter plot shown that images with high sharpness happen at the rest state.}
\label{fig:AccVSS2_obs}
\end{figure}

This work is motivated by the design of a context awareness framework to enable complex control strategies and adaptation in a prosthetic leg. The context awareness that we consider incorporates the following components:
\begin{itemize}
\item{\bf A Vision System.} Composed of a camera and a processing unit capable of performing tasks such as feature extraction, Simultaneous Localization and Mapping (SLAM), and classification.
\item{\bf An Inertial System.} Composed of Inertial Measurement Units (IMUs) responsible for providing data such as acceleration and angular rate. This sensing modality can be fused to provide information to help determine the terrain slope, track the state of the prosthesis, and track the gait cycle of the user.
\end{itemize}

\begin{figure}[h]

\begin{subfigure}[t]{\linewidth}
\centering
\includegraphics[width=\linewidth]{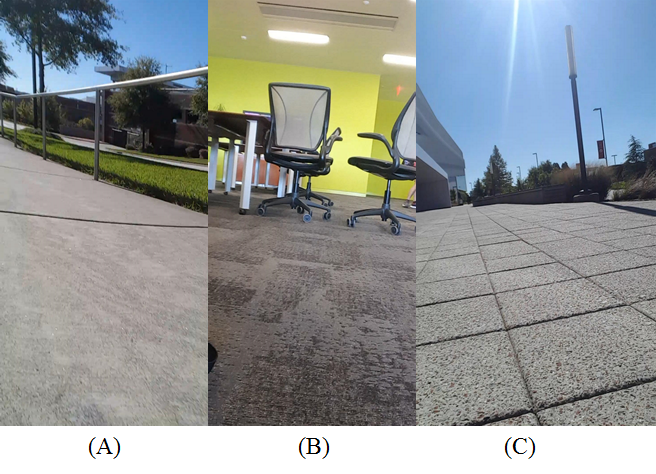}
\end{subfigure}
~
\begin{subfigure}[t]{\linewidth}
\centering
\includegraphics[width=\linewidth]{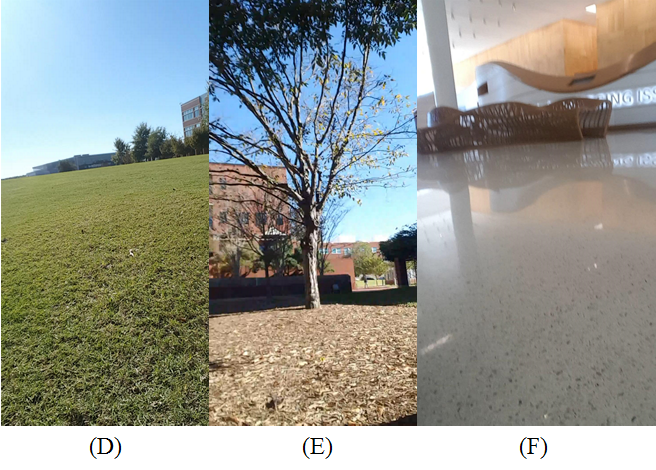}
\end{subfigure} 

\caption{Samples of the different type of terrains observed from the camera on the leg. (A) Asphalt, (B) Carpet, (C) Cobblestone, (D) Grass, (E) Mulch and (F) Tile.} 
\label{fig:Terrain_Samples}

\end{figure}

The objective of the system is to provide some context awareness for the prosthesis. Hence, the images must be captured and processed through some recognition pipeline. Since some of the frames in the video stream can suffer from motion blur, we must consider a key-frame selection approach. As in \cite{Anantrasirichai2015}, we use sharpness as a measurement of image quality. The rest position of a gait cycle takes place from the moment the foot first makes contact with the ground until the moment when a toe leaves the floor \cite{Collin2003}. During this time, the acceleration of the foot is nearly constant, allowing the use of the accelerometer to trigger the capture of a low blur image.

The run time in the processing unit can be reduced by choosing which frames should be used before doing the computationally intensive tasks. Selection of low blur images can be achieved by identifying when the foot is in rest position using an IMU (i.e.,  the main acceleration measured is just due to the constant effect of gravity).Figure \ref{fig:AccVSS2_obs} shows the relationship between acceleration and sharpness.


To compare the computational capabilities of the Raspberry Pi 3 and NVIDIA Jetson TX2, we implemented a standard image classification pipeline to run on both platforms. Our image classification pipeline for speed evaluation includes two steps: (1) feature extraction, and (2) classification. Feature extraction was done with LBP (Local Binary Pattern) \cite{scikit-image}, while the utilized classifier is the Random Forest (RF) \cite{scikit-learn}.

The data used for evaluation has been collected using a Samsung S6 cellphone, attached to the shin of the subjects. The protocol to collect data consisted of recording videos from two different subjects, while walking at a regular pace on the North Carolina State University (NCSU) campus for about 20 minutes. The subjects walked over different types of terrains for about 3 to 4 minutes. The dataset was narrowed down to the images with low blur and ended up consisting of 566 images of asphalt, 648 images of carpet, 1041 images of cobblestone, 984 images of grass, 168 images of mulch and 585 of tile (see Figure \ref{fig:Terrain_Samples} for samples), which reflect the terrains the subjects walked over.

\section{PLATFORM EVALUATION}
\label{sec:plateval}

We now describe how the evaluation was performed. First, we provide details about how the algorithms described on Section \ref{sec:motapp} were implemented followed by a description of how the tests were designed and executed.

\subsection{Implementation Details}
For LBP feature extraction, we set the number of circularly symmetric neighbor points to be 24; the radius of circle to be 8 pixels; and to improve the rotation invariance with uniform patterns. For the random forest classifier, we set the number of trees in the forest to be 100. The two platforms have Python 2.7 \cite{python27} and OpenCV 3.3 \cite{OpenCV33} libraries installed. The LBP method was obtained from skimage 0.13.1 \cite{scikit-image}, while the RF was obtained from sklearn 0.19.1 \cite{scikit-learn}. 

\subsection{CPU Load and Memory Usage}
\label{sec:CPUload}
The dataset described in Section \ref{sec:motapp} was utilized to train the RF classifier using the LBP description of the images to determine the terrain. The test set was composed of 1000 images. The training was done on a regular desktop while the testing was executed on both platforms.

For testing, we computed the average run time for each execution. The subset of 1000 images from the dataset was evaluated 10 times. Also, we monitored the CPU load and memory usage percentage by sending system commands every 100 $ms$ while the testing of the data was being done. At the end of each evaluation of the dataset, we calculated the average CPU load and memory usage for that particular experiment. This generated 10 values of average CPU load and average memory usage. Finally, the results out of the 10 executions were averaged, providing an average of the run time, CPU load and memory usage. In order to have a notion of runtime in terms of frames per seconds, we took the final average runtime value and divided by 1000. This procedure was then repeated for each of the image resolutions: 32$\times$32, 64$\times$64, 128$\times$128, 256$\times$256, 512$\times$512 and 1024$\times$1024.

Tables \ref{tab:cpumem} and \ref{tab:memusage} show the comparison of CPU load and memory usage between the two platforms. Both values are approximately constant across the image resolutions, with a slightly increase of used resources for higher resolutions such as 512$\times$512 and 1024$\times$1024. The almost constant memory usage may be due to the small size of the images compared to the resources needed to load the application and corresponding libraries.

\begin{table}[h]
\centering
\caption{CPU Load [\%]}
\label{tab:cpumem}
\begin{tabular}{l|c|c|c|c|c|c|}
\cline{2-7}
\multicolumn{1}{c|}{}                & \multicolumn{6}{c|}{Resolution}                                                                                                                                            \\ \cline{2-7} 
                                     & 32                         & 64                         & 128                        & 256                        & 512                        & 1024                      \\ \hline
\multicolumn{1}{|l|}{Raspberry Pi 3} & 24.77                      & 24.99                      & 25.00                      & 25.01                      & 25.08                      & 25.07                      \\ \hline
\multicolumn{1}{|l|}{Jetson TX2}     & \multicolumn{1}{l|}{16.57} & \multicolumn{1}{l|}{16.54} & \multicolumn{1}{l|}{16.69} & \multicolumn{1}{l|}{16.65} & \multicolumn{1}{l|}{18.04} & \multicolumn{1}{l|}{25.31} \\ \hline
\end{tabular}
\end{table}

\begin{table}[h]
\centering
\caption{Memory RAM Usage [\%]}
\label{tab:memusage}
\begin{tabular}{l|c|c|c|c|c|c|}
\cline{2-7}
\multicolumn{1}{c|}{}                & \multicolumn{6}{c|}{Resolution}                                                                                                                                       \\ \cline{2-7} 
                                     & 32                        & 64                        & 128                       & 256                       & 512                       & 1024                      \\ \hline
\multicolumn{1}{|l|}{Raspberry Pi 3} & 8.72                      & 8.69                      & 8.65                      & 8.75                      & 9.17                      & 10.78                     \\ \hline
\multicolumn{1}{|l|}{Jetson TX2}     & \multicolumn{1}{l|}{1.24} & \multicolumn{1}{l|}{1.24} & \multicolumn{1}{l|}{1.27} & \multicolumn{1}{l|}{1.25} & \multicolumn{1}{l|}{1.30} & \multicolumn{1}{l|}{1.54} \\ \hline
\end{tabular}
\end{table}

\begin{figure}[!b]

\centering
\includegraphics[width=\linewidth]{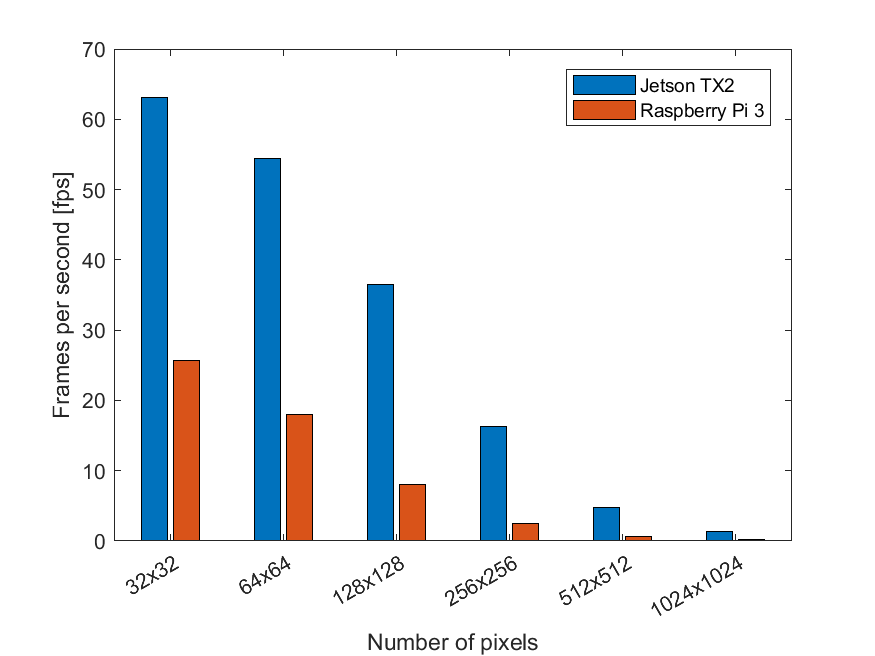}

\caption{Run time comparison}
\label{fig:runtime}

\end{figure}

Figure \ref{fig:runtime} shows the comparison of runtime between the two platforms. For low resolutions such as 32$\times$32,  64$\times$64, 128$\times$128, one can see that the Jetson TX2 rates superior to video are achieved capped at 36$fps$. However, for resolutions higher than 128$\times$128, it would be necessary to select specific frames. For example, if the Jetson TX2 is embedded in a prosthetic leg, one can trigger an image capture when the leg is nearly static, which could be identified with the help of an IMU (as we described in Section \ref{sec:motapp}). Using the 512$\times$512 images, we would get 4.7 frames per second for the processing, while with 1024$\times$1024 (which is similar in terms of number of pixels to 720$\times$576 and 1280$\times$720 image resolutions), we could process 1.3 frames per second. Therefore, for resolutions up to 512$\times$512, utilizing the Jetson board could enable embedded implementation of methods such as surface reconstruction and provide the terrain slope at which the user are going to step on, consequently, allowing a more adaptive control of the prosthetic, section \ref{sec:accuracyassess} shows an application example with image classification. Depending on the application, the same could be done with the Raspberry Pi 3, however, at much smaller resolutions (capped about 128$\times$128 pixels for instance), resolution higher than this are harder to work with since the rates obtained for 512$\times$512 was 0.6 $fps$ and 0.15 $fps$ for 1024$\times$1024.

\subsection{Accuracy across different image resolutions}
\label{sec:accuracyassess}
\begin{figure}

\centering
\includegraphics[width=\linewidth]{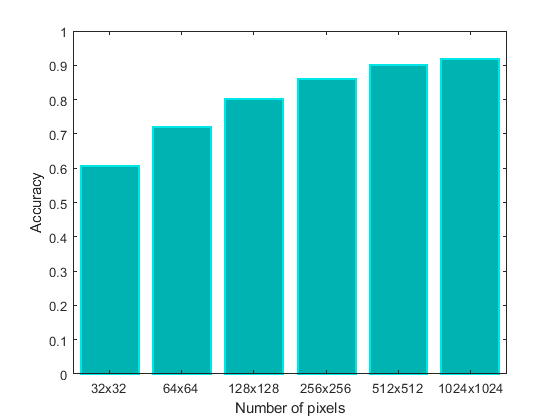}

\caption{Classification accuracy comparison} 
\label{fig:accuracy}

\end{figure}

As it was described on section \ref{sec:CPUload}, the dataset was randomly divided into training and testing sets. Specifically, the dataset was randomly divided into training $(70\%)$ and testing $(30\%)$ sets. In order to study the influence of the image resolutions on the classification accuracy, we cropped our original dataset images to 1024$\times$1024. The cropping was done in the middle of the image. Then, the cropped set was resized to each of the evaluated resolutions: 32$\times$32, 64$\times$64, 128$\times$128, 256$\times$256 and 512$\times$512. A specific RF classifier was trained for each image resolution. The testing was done in order to assess how much the rate decrease would degrade the classification accuracy. As shown in Figure \ref{fig:accuracy}, there is no significant loss of accuracy by reducing the frame rate from 1024$\times$1024 to 512$\times$512. The next step below to 256$\times$256 would degrade the previous accuracy 4\% down and the next step down to 128$\times$128 would degrade extra 6\%. As an illustration, the confusion matrix respective to 128$\times$128, 256$\times$256 and 512$\times$512 are shown by the tables \ref{tab:confumatrix128} to \ref{tab:confumatrix512}. The matrices show us how many figures were correctly identified.

\begin{table}[h!]
\centering
\caption{Confusion Matrix for 128$\times$128 pixels}
\label{tab:confumatrix128}
\begin{tabular}{l|l|l|l|l|l|l|}
\cline{2-7}
\multicolumn{1}{c|}{}             & \multicolumn{6}{c|}{Predicted}                                                                                                                            \\ \hline
\multicolumn{1}{|l|}{Known}       & \multicolumn{1}{c|}{As} & \multicolumn{1}{c|}{Ca} & \multicolumn{1}{c|}{Co} & \multicolumn{1}{c|}{Gr} & \multicolumn{1}{c|}{Mu} & \multicolumn{1}{c|}{Ti} \\ \hline
\multicolumn{1}{|l|}{\textbf{As}phalt}     & 120                    & 6                    & 30                    & 1                    & 0                    & 5                    \\
\multicolumn{1}{|l|}{\textbf{Ca}rpet}      & 2                    & 151                    & 5                    & 12                   & 1                    & 15                    \\
\multicolumn{1}{|l|}{\textbf{Co}bblestone} & 11                    & 8                    & 280                    & 20                   & 0                    & 4                    \\
\multicolumn{1}{|l|}{\textbf{Gr}ass}       & 2                    & 16                    & 24                   & 250                    & 4                    & 2                    \\
\multicolumn{1}{|l|}{\textbf{Mu}lch}       & 2                    & 2                    & 12                   & 6                    & 24                    & 0                    \\
\multicolumn{1}{|l|}{\textbf{Ti}le}        & 3                    & 21                    & 11                   & 10                  & 0                    & 142                    \\ \hline
\end{tabular}
\end{table}

\begin{table}[h!]
\centering
\caption{Confusion Matrix for 256$\times$256 pixels}
\label{tab:confumatrix256}
\begin{tabular}{l|l|l|l|l|l|l|}
\cline{2-7}
\multicolumn{1}{c|}{}             & \multicolumn{6}{c|}{Predicted}                                                                                                                            \\ \hline
\multicolumn{1}{|l|}{Known}       & \multicolumn{1}{c|}{As} & \multicolumn{1}{c|}{Ca} & \multicolumn{1}{c|}{Co} & \multicolumn{1}{c|}{Gr} & \multicolumn{1}{c|}{Mu} & \multicolumn{1}{c|}{Ti} \\ \hline
\multicolumn{1}{|l|}{\textbf{As}phalt}     & 135                    & 2                    & 24                    & 9                    & 0                    & 5                    \\
\multicolumn{1}{|l|}{\textbf{Ca}rpet}      & 2                    & 154                    & 0                    & 6                    & 0                    & 14                    \\
\multicolumn{1}{|l|}{\textbf{Co}bblestone} & 4                    & 5                    & 273                    & 7                    & 0                    & 4                    \\
\multicolumn{1}{|l|}{\textbf{Gr}ass}       & 3                    & 11                    & 9                    & 301                    & 3                    & 0                    \\
\multicolumn{1}{|l|}{\textbf{Mu}lch}       & 1                    & 1                    & 3                   & 11                    & 29                    & 2                    \\
\multicolumn{1}{|l|}{\textbf{Ti}le}        & 3                    & 34                    & 3                    & 1                    & 1                    & 142                    \\ \hline
\end{tabular}
\end{table}

\begin{table}[h!]
\centering
\caption{Confusion Matrix for 512$\times$512 pixels}
\label{tab:confumatrix512}
\begin{tabular}{l|l|l|l|l|l|l|}
\cline{2-7}
\multicolumn{1}{c|}{}             & \multicolumn{6}{c|}{Predicted}                                                                                                                            \\ \hline
\multicolumn{1}{|l|}{Known}       & \multicolumn{1}{c|}{As} & \multicolumn{1}{c|}{Ca} & \multicolumn{1}{c|}{Co} & \multicolumn{1}{c|}{Gr} & \multicolumn{1}{c|}{Mu} & \multicolumn{1}{c|}{Ti} \\ \hline
\multicolumn{1}{|l|}{\textbf{As}phalt}     & 141                  & 2                    & 14                   & 8                    & 0                    & 3                    \\
\multicolumn{1}{|l|}{\textbf{Ca}rpet}      & 2                    & 174                  & 1                    & 2                    & 0                    & 9                    \\
\multicolumn{1}{|l|}{\textbf{Co}bblestone} & 9                    & 5                    & 281                  & 9                    & 0                    & 3                    \\
\multicolumn{1}{|l|}{\textbf{Gr}ass}       & 5                    & 5                    & 8                    & 286                  & 2                    & 1                    \\
\multicolumn{1}{|l|}{\textbf{Mu}lch}       & 3                    & 1                    & 1                   & 12                    & 32                    & 0                    \\
\multicolumn{1}{|l|}{\textbf{Ti}le}        & 0                    & 12                   & 1                    & 0                    & 0                    & 170                    \\ \hline
\end{tabular}
\end{table}

\section{PROPOSED ARCHITECTURE}
\label{sec:proparch}
Figure \ref{fig:SystemOverview} illustrates our proposed architecture for enabling visual sensing on a lower limb prosthesis. Based on our analysis from the previous sections, we can propose the use of a hierarchical architecture for handling the real-time requirements and minimize power requirements. 

\paragraph{Real-Time}
Our system architecture would include the vision sensing unit (VSU) described in section \ref{sec:motapp} as well as a camera whose frames would be processed by a Jetson TX2. The processing  of IMU data and EMG signals required for control could be handled by an embedded platform that can ensure real-time processing (e.g., with capabilities similar to the Raspberry Pi 3) since those tasks are not computationally demanding. This would free up resources utilized by the VSU, enabling it to perform adaption and even learning-related tasks. The IMU would provide the signals to help in the estimation of terrain slope as well as identify the gait phases, enabling the recognition of nearly static movements and triggering the capture of low blur images on the VSU. Information coming from the VSU (such as terrain identification or descriptors such as inclination) can then be used as context-awareness inputs to the controller. It is not desirable to have all computation taking place in the Jetson since the priority of the processes are handled by the OS, which could affect the real-time feedback required for safety of the device. From our analysis, we observe that it would be possible to pick a frame during a single period of a gait in order to recognize terrain. This means that we would need to perform all visual processing within a period of length dependent on the walking pace (typical values are about 1s). Such strategy enables the use of more complex inference within the Jetson (e.g., using deep learning pipelines).

\paragraph{Power Consumption.} Compared to the power needed to support the actuator (e.g., the emPOWER ankle from the BionX Medical Technology has maximum power ~700W, and 45Whr energy), the power needed for on-board computation (e.g., ~7.5 W for NVIDIA Jetson TX2) is small. However, since the computing needs to be persistent, this can cause large power requirements. Therefore, the inclusion of the Raspberry PI 3 would reduce the processing bandwidth required of the VSU, thus reducing the power consumption of the VSU as well.

\section{CONCLUSIONS}

The comparison between the Raspberry Pi 3 and Jetson TX2 allowed us to envision the application of CV techniques to enhance context awareness for a robotic prosthetic leg. The Jetson board could be used for ``visual intensive'' processing, while the Raspberry Pi (or similar real-time platform) could be used together with the Jetson, depending on the application. However, in terms of visual terrain recognition, the CPU load and memory usage of the Jetson were not high, bringing evidence that the Raspberry Pi could be dedicated to tasks such as sensor fusion of data streams and real-time control. Additionally, the similarity of classification accuracy observed between the images of sizes 512$\times$512 and 1024$\times$1024 reveals that working with reduced image sizes may allow similar results to higher quality images with the benefit of saving processing bandwidth.

A further consideration for the use of visual sensors on lower limb prosthesis is privacy. The use of key-frame selection using the IMU removes the need for capturing video which reduces the amount of utilized resources on the Jetson, whereas a solution using video would have to autonomously blur any places where a face may be present.


\section*{ACKNOWLEDGMENT}
We thank Jean~P. Diaz for his help on the data collection and suggestions on the paper. We also thank Jeremy Cole for his help on proofreading of the manuscript.

\bibliographystyle{ieeetr}
\bibliography{prosthetic.bib}

\end{document}